# Ted Geballe:  A Lifetime of Contributions To Superconductivity

G. R. Stewart

Department of Physics, University of Florida, Gainesville, FL 32611

The editors' have dedicated this special issue on superconducting materials "to Ted Geballe in honor of his numerous seminal contributions to the field of superconducting materials over more than 60 years, on the year of his 95th birthday." Ted has made so many contributions to the study of superconductivity - from his co-discovery with Matthias et al. of 18 K superconductivity in $Nb_3Sn$ [1] to his recent research [2] on overdoped cuprates, $T_c$~90 K, with Marezio et al. – that to fairly summarize this body of work would require many more pages than allotted here.  Put another way:  the discovery of $Nb_3Sn$ is #8 on Ted's publication list, and the investigation of overdoped cuprates is #486.  Thus, the reader is invited to let the articles herein on the different superconducting materials classes and their many references to Ted's work (my article on the A15's references six separate papers of his) speak to the *detailed* specifics of Ted's contributions.  Here, as an executive summary, are just a few highlights of his research in superconductivity, leavened with some anecdotes from when I had the good fortune to be his graduate student from 1971-1975, and ending with some of Ted's general insights and words of wisdom.

Before beginning, it should be stressed that to discuss the part of Ted's activities related to superconductivity is to discuss only a fraction of the total package.  Ted is interested not only in the full range of solid state physics, but also technology, art, literature, and many other areas of human creativity.  In Ted's early years at Bell Labs, for example, he also worked extensively on semiconductors – 7 of his first 25 publications (on Si, Ge, and InSb) have been cited a total of over 1800 times.  One of these, his study [3] of isotopic effects on thermal conductivity in Ge, can be found as a figure in chapter 5 in Kittel Introduction to Solid State Physics.  Ted's students and postdocs were always involved in a number of areas of research at a given time, with superconductivity just one of several main foci.

As an example of his technological advances that were important in superconductivity research, Ted's original work in 1955 and 1957 ("Germanium resistance thermometers suitable for low-temperature calorimetry" [4],) in doped Germanium had as its direct result the secondary standard used for low temperature thermometry the world over since it was commercialized in the 1960's.  Later, in 1972, Ted's group at Stanford developed [5] the time constant method of low temperature calorimetry that, with an improvement in thermometry in 1975, directly led to the automated calorimeter sold by the 100's today by Quantum Design and in use all over the world.  Of course, both these advances have far broader application that just characterizing superconducting materials.

A few (idiosyncratic) highlights of Ted's research in superconductivity:  All of us learned that the BCS theory predicts an isotope effect, that the mass affects the superconducting transition temperature via $T_c \propto M^{-1/2}$. We also learned that there are exceptions, understood from the Morel-Anderson theory (1962) for the renormalized Coulomb interaction µ*. It was Ted's discovery in 1961 [6] that $T_c$ for Ru was approximately independent of the isotopic mass that was the first known exception. Another example – from Ted's more 'thinking outside the box' side – that reflects Ted's constant search for the unusual, for the breakthrough idea, started during my grad student days. Sasha Rusakov visited Ted's theory colleague Walt Harrison, also in the Applied Physics Department, with ideas about CuCl possibly being a superconductor. Several years later Ted, together with Paul Chu, Rusakov, and others published [7] a paper reporting a 7% diamagnetic signal in the ac susceptibility at 240 K. This CuCl work serves as an example of one of Ted's lasting themes – the search for a novel high temperature superconducting phase hidden as a minority phase in the parent material.

This CuCl example segues over into my discussion of Ted's general insights. An overriding philosophy of Ted's in his search for new materials was to look, at least part of the time, off the beaten path: "enhanced superconductivity can be caused by an unrecognized inhomogeneity." Another theme that I learned from Ted (although as a grad student I didn't know any other way and couldn't imagine that it could ever be otherwise), and took to heart for my own work ever since: it is really *much* better to have the sample making embedded in the same group doing the characterization. During my 4 years as a grad student, John Gavaler at Westinghouse R & D discovered a new high temperature superconductor in 1973: ~23 K superconductivity in sputtered A15 $Nb_3Ge$. This caused great excitement in the whole superconductivity community. Ted's group, under the direction of Bob Hammond, began to make high quality e-beam A15 films and used them to study the underlying physics. Several high quality scientific papers came out every year for a number of years starting in 1976, as well as a number of PhD theses, based on these films. Being able to literally go next door and make this or that modification in the samples was a tremendous advantage in the research.

Finally, a few of my favorite quotes from Ted: "Theories can come and go, but accurate experimental results can be forever." "It's good to talk to theorists – but even if they tell you your experiments won't do what you think, it doesn't mean you shouldn't try." Finally, somewhat paraphrased after all the years when he said it as part of my training, "Even though you may think of yourself as an experimentalist, since you use theory both to explain your data and design future experiments – you're also part theorist."

Certainly Ted's command of, and interest in, theory, technological innovation, and experimental techniques has given his work in superconductivity and superconducting materials an enormous impact.

From all of those with whom Ted has interacted both professionally and personally over more than six decades and who have benefitted from this work: 'Thanks Ted!'

**Acknowledgements:** Work at Florida performed under the auspices of the US Department of Energy, Office of Basic Energy Sciences, contract no. DE-FG02-86ER45268.